\documentclass[]{aa}
\usepackage{epsfig}

\graphicspath{{zzzFigures/}}

\catcode`\@=11
\def\gsim{\ifmmode{\mathrel{\mathpalette\@versim>}}
    \else{$\mathrel{\mathpalette\@versim>}$}\fi}
\def\lsim{\ifmmode{\mathrel{\mathpalette\@versim<}}
    \else{$\mathrel{\mathpalette\@versim<}$}\fi}
\def\@versim#1#2{\lower 2.9truept \vbox{\baselineskip 0pt \lineskip
    0.5truept \ialign{$\m@th#1\hfil##\hfil$\crcr#2\crcr\sim\crcr}}}
\catcode`\@=12
\def\Re{R_{\rm e}}
\def\Reo{R_{\rm e}^0}
\def\ellef{\ell_{\rm e}}
\def\rc{R_{\rm c}}
\def\rap{R_{\rm a}}
\def\rBH{r_{\rm BH}}

\def\Ie{I_{\rm e}}

\def\qt{q(\theta)}

\def\Lz{L_z}

\def\los{{\it los }}
\def\xv{{\bf x}}

\def\xp{x\prime}
\def\yp{y\prime}
\def\zp{z\prime}

\def\nv{{\bf n}}

\def\vv{{\bf v}}

\def\ttv{{\rm v}}

\def\tvphi{\ttv_{\varphi}}
\def\tvphim{\overline{\ttv}_{\varphi}}
\def\tvphis{\tvphi^2}
\def\tvphism{\overline{\tvphis}}

\def\vcz{v_z}

\def\vcR{v_R}
\def\vcphi{v_{\varphi}}
\def\vn{v_{\rm n}}
\def\vp{v_{\rm p}}
\def\Vp{V_{\rm p}}
\def\Vps{\Vp^2}

\def\sigc{\sigma_0}
\def\sigR{\sigma_R}
\def\sigRs{\sigR^2}
\def\sigz{\sigma_z}
\def\sigzs{\sigz^2}

\def\sigphi{\sigma_{\varphi}}
\def\sigphis{\sigphi^2}

\def\sigp{\sigma_{\rm p}}
\def\sigps{\sigp^2}
\def\sigob{\sigma_{\rm los}}

\def\sigobs{\sigob^2}

\def\sigap{\sigma_{\rm a}}

\def\onesigint{\sigma_{\rm int}}

\def\rms{{\it rms}}
\def\ml{\Upsilon_\ast}
\def\Kvir{K_{\rm V}}
\def\MBH{M_{\rm BH}}

\def\lc03{\mbox{LC}}
\def\cb05{\mbox{CB}}
\def\dg{\mbox{DG}}

\begin{document}
   \title{The contribution of rotational velocity to the FP of
     elliptical galaxies}


   \author{A. Riciputi\inst{1}, B. Lanzoni\inst{2}, S. B\`{o}noli\inst{3}  \and 
           L. Ciotti\inst{1}}

   \offprints{andrea.riciputi@bo.astro.it}

   \institute{Dipartimento di Astronomia, 
              Universit\`a di Bologna, via Ranzani 1,
              I-40127 Bologna, Italy
         \and
              INAF - Osservatorio Astronomico di Bologna, via Ranzani 1,
              I-40127 Bologna, Italy
         \and
              Department of Astronomy \& Astrophysics, University of
              Toronto, 60 St.\ George Street, Toronto, ON, M5S 3H8, Canada}

   \date{Received 8 June 2005/ Accepted 28 July 2005}
   
   \abstract{The contribution of ordered rotation to the observed tilt
     and thickness of the Fundamental Plane of elliptical galaxies is
     studied by means of oblate, two-integrals cuspy galaxy models with
     adjustable flattening, variable amount of ordered rotational
     support, and possible presence of a dark matter halo and of a
     central super-massive black hole. We find that, when restricting
     the measure of the velocity dispersion to the central galactic
     regions, rotation has a negligible effect, and so cannot be
     responsible of the observed tilt. However, streaming velocity
     effect can be significant when observing small and rotationally
     supported galaxies through large (relative) aperture (as for
     example in Fundamental Plane studies at high redshift), and can
     lead to unrealistically low mass-to-light ratios. The effect of a
     central supermassive black hole on the kinematical fields, and the
     models position in the $v/\sigma$-ellipticity plane are also
     discussed.

}

\authorrunning{Riciputi et al.}
\titlerunning{Rotational effects on the FP thickness}
\maketitle

\section{Introduction}

In the observational three-dimensional space of central velocity
dispersion $\sigc$, (circularized) effective radius $\Re$, and mean
surface brightness within the effective radius $\Ie = L / (2 \pi
\Re^2)$ (where $L$ is the total galaxy luminosity), early-type
galaxies approximately locate on a plane, called the Fundamental Plane
(hereafter FP; Dressler et al. 1987; Djorgovski \& Davis 1987), and
represented by the best-fit relation:
\begin{equation}
\log\Re =a\log\sigc +b\log\Ie +c.
\label{eq:FP}
\end{equation}
The coefficients $a$, $b$, and $c$ depend slightly on the considered
photometric band (e.g., Pahre, de Carvalho, \& Djorgovski, 1998;
Scodeggio et al. 1998): for example, by measuring $\Re$ in kpc, $\sigc$
in km/s, and $\Ie$ in $L_\odot/$pc$^2$, reported values in the Gunn r
band are $a =1.24\pm 0.07$, $b =-0.82\pm 0.02$, $c =0.182$\footnote{This
  value of $c$ refers to the Coma cluster and to $H_0=50$ km s$^{-1}$
  Mpc$^{-1}$. Note that $\sigc$ is usually corrected to a circular
  aperture with diameter $1.19\,h^{-1}$ kpc, corresponding to a radial
  range $\sim 0.05\,\Re$--$\Re$ for $h=0.5$, and for typical values of
  $\Re$ (e.g., J{\o}rgensen, Franx, \& Kj{\ae}rgaard 1995).}
(J{\o}rgensen, Franx, \& Kj{\ae}rgaard 1996, hereafter JFK).  One of the
most striking observational properties of the FP is its small and nearly
constant thickness: the distribution of $\log\Re$ around the best-fit FP
(at fixed $\sigc$ and $\Ie$) has a measured $\rms$ (hereafter
$\onesigint$) corresponding to a scatter in $\Re$ $\sim 15\% \div 20\%$
(after correction for measurement errors, see, e.g., Faber et al. 1987;
JFK).

In addition, for a stationary stellar system the scalar virial theorem
can be written as
\begin{equation}
{G\ml L\over\Re}=\Kvir\sigc^2,
\label{eq:VT}
\end{equation}
where $\ml$ is the {\it stellar} mass-to-light ratio in the photometric
band used for the determination of $L$ and $\Re$, while the coefficient
$\Kvir$ takes into account projection effects, the galaxy stellar
density distribution, the stellar orbital distribution (such as velocity
dispersion anisotropy and rotational support), and the effects related
to the presence of dark matter.  Equations (\ref{eq:FP}) and
(\ref{eq:VT}) imply that in real galaxies, no matter how complex their
structure is, $\ml/\Kvir$ is a well-defined function of any two of the
three observables $(L,\Re,\sigc)$ and this dependence is commonly
referred as the ``FP tilt''. In other words, the FP tilt indicates that
{\it structural/dynamical} ($\Kvir$) and {\it stellar population}
($\ml$) properties in real galaxies are strictly connected, possibly as
a consequence of their formation process. Overall, the interpretation of
the FP cannot be limited to the study of its tilt only, but requires to
take consistently into account also its thinness (as done, for example,
in the ``orthogonal exploration'' approach, see Renzini \& Ciotti 1993,
Ciotti, Lanzoni \& Renzini 1996, Ciotti \& Lanzoni 1997).  A step
forward in the study of the structural and dynamical implications of the
tilt and thinness of the FP was made by Bertin, Ciotti \& Del Principe
(2002), who introduced, although limited to spherical models, the more
general Monte-Carlo approach. The Monte-Carlo approach was successively
applied by Lanzoni \& Ciotti (2003, hereafter \lc03) to the
investigation of projection effects on the FP thickness, by adopting
fully analytical axisymmetric one-component Ferrers galaxy models
(Ferrers 1877).  In particular, \lc03 found that while projection
effects do contribute to the observed FP scatter, nonetheless the FP
\emph{physical scatter} (as determined by variations of the physical
properties from galaxy to galaxy), sums up to 90\% of the intrinsic FP
scatter.  In addition, \lc03 found that the contribution of ordered
streaming motions to the observed velocity dispersion is negligible when
small/medium apertures ($\lsim\,\Re$) are used for the spectroscopic
observations. Unfortunately, while fully reliable about projection
effects on $\Re$ (which are independent of the specific homeoidal
density profile adopted), the central regions of Ferrers models are
unrealistically flat, so that doubts can be risen about the use of their
central velocity dispersion to estimate streaming velocity effects.
Moreover, what is the additional effect of a central super-massive black
hole (SMBH) or of a dark matter halo? An attempt to answer the questions
above was made by Ciotti \& Bertin (2005, hereafter \cb05{}) by using
fully analytical cuspy galaxy models obtained from a homeoidal
expansion, qualitatively confirming the results of \lc03{}. Here, in
order to extend the results of \lc03{} and \cb05{}, we numerically solve
the Jeans and projected dynamics equations for more realistic, oblate
galaxy models with a central cusp, and variable amount of internal
streaming velocity, allowing for the presence of a central SMBH and a
dark matter halo.

The paper is organized as follows. In Sect.~\ref{sec:model-prop} we
derive the relevant intrinsic and projected properties of the adopted
models, while in Sects.~\ref{sec:results}, \ref{sec:fp-thickness}
and~\ref{sec:vsigma-epsilon-plane} we present the main results of our
analysis together with some observationally related consequences, such
as the model position in the $v/\sigma$-ellipticity plane. In
Sect.~\ref{sec:disc-concl} we summarize the results, while in the
Appendix we present the fully analytical model adopted to test the
numerical code.

\section{The models}
\label{sec:model-prop}
\subsection{3D quantities}
\label{sec:3d-quantities}

At variance with \lc03{}, who adopted the centrally flat and spatially
truncated Ferrers model, we now use cuspy oblate galaxy models with homeoidal
density distribution, belonging to the family of the so-called
$\gamma$-models (Dehnen 1993; Dehnen \& Gerhard 1994, hereafter \dg;
Tremaine et al. 1994; Qian et al. 1995).  Their density profile is
\begin{equation}
  \rho_{\ast} = \frac{M_\ast}{4\pi \, \rc^3 \, q} \, 
  \frac{(3 - \gamma)}{m^\gamma \, (1+m)^{4-\gamma}}, ~~~~ (0\le\gamma <3),
  \label{eq:rho}
\end{equation}
where $M_\ast$ is the total mass, $\rc$ the characteristic scale, and in
cylindrical coordinates\footnote{The $(R,\varphi, z)$ coordinates are
  related to the natural Cartesian coordinates by the relations
  $R=\sqrt{x^2 + y^2}$, $\cos\varphi=x/R$, $\sin\varphi=y/R$. From now
  on, the symbol ``$\sim$'' over a coordinate will indicate
  normalization to $\rc$.}  $m \equiv \sqrt{\tilde{R}^2 +
  \tilde{z}^2/q^2}$, with $\tilde{R} \equiv R / \rc$ and $\tilde{z}
\equiv z / \rc$; the parameter $0 < q \leq 1$ measures the intrinsic
model flattening. Note that these models for $\gamma \simeq 0.75$ and
$\gamma \simeq 1$ provide a good representation (in the central regions)
of the de Vaucouleurs (1948) and of the Navarro, Frenk, \& White (1997)
profiles, respectively. In two-component models the dark matter halo is
also described by a density profile as in Eq.~(\ref{eq:rho}), in general
with a different total mass, scale-radius, density slope and flattening;
a SMBH of mass $\MBH$ can be also added at the model center. In all
cases, the stellar mass-to-light ratio $\ml$ is assumed to be constant
within each model.

We assume that the density profile in Eq.~(\ref{eq:rho}) is supported by
a two-integrals distribution function $f(E,\Lz)$\footnote{Note that the
  important issue of the models phase-space consistency is beyond the
  tasks of this work (see Qian et al.\ 1995). We use symbol $\vv$ for
  the velocity in the phase space, while $\vec v(\xv)\equiv
  \overline{\vv}$ is the {\it streaming} velocity.  In general, a bar
  over a quantity means average over phase-space velocities.}  (where
$E$ and $\Lz$ are the energy and the $z$-component of the angular
momentum of stars per unit mass, respectively), and the Jeans equations
are
\begin{eqnarray}
    \frac{\partial\rho_\ast \sigRs}{\partial z} &=& -\rho_\ast {\partial\phi\over\partial
      z}, \label{eq:jeans1}\\
    {\partial\rho_\ast\sigRs\over\partial R} &-& {\rho_\ast (\tvphism -\sigRs)\over R} = 
    -\rho_\ast {\partial\phi\over\partial R}, \label{eq:jeans2}
\end{eqnarray}
where $\phi$ is the total gravitational potential, $\vcR = \vcz = 0$,
the off-diagonal elements of the velocity dispersion tensor vanish, and
$\sigRs = \sigzs$ (see, e.g., Binney \& Tremaine 1987, hereafter BT).

As in \lc03{}, $\tvphism$ is splitted into streaming motion $\vcphi^2
\equiv \tvphim^2$ and azimuthal velocity dispersion
$\sigphis\equiv\overline{(\tvphi -\vcphi)^2}=\tvphism -\vcphi^2$ with
the Satoh (1980) $k$-decomposition:
\begin{equation}
\vcphi^2 = k^2(\tvphism -\sigRs),
\label{eq:vphi}
\end{equation}
and
\begin{equation}
\sigphis =\sigRs + (1-k^2) (\tvphism - \sigRs),
\label{eq:sigphi}
\end{equation}
with $0\le k\le 1$.  For $k=0$ no ordered motions are present, and the
velocity dispersion tensor is maximally tangentially anisotropic, while
for $k=1$ the velocity dispersion tensor is isotropic, and the galaxy
flattening is due to azimuthal streaming velocity (the so-called
``isotropic rotator'' case).

As well known, the gravitational potential of the density distribution
in Eq.~(\ref{eq:rho}) cannot be expressed in terms of elementary
functions, and so we compute it by using an expansion in orthogonal
function: Eqs.~(\ref{eq:jeans1}) and~(\ref{eq:jeans2}) are then
integrated numerically (Sect.~\ref{sec:numcode}).

\subsection{Projected quantities}
\label{sec:projected-quantities}
In order to project the model properties on the plane of the sky (the
projection plane), we define the observer system as a Cartesian
coordinate system $(\xp, \yp, \zp)$ with the line-of-sight (\emph{los})
parallel to the $\zp$ axis, and with the $\xp$ axis coincident with the
$x$ axis of the natural Cartesian system introduced in
Sect.~\ref{sec:3d-quantities}.  The angle between $z$ and $\zp$ is
$0\le\theta\le\pi /2$: $\theta=0$ corresponds to the face-on view of the
galaxy, while $\theta=\pi/2$ to the edge-on view.  With this choice, the
projection plane is $(\xp,\yp)$ and the \los direction in the natural
coordinate system is given by $\nv =(0,-\sin\theta
,\cos\theta)$\footnote{The \los vector points \emph{toward} the
  observer, and so \emph{positive} velocities correspond to a
  \emph{blue--shift}; the general framework for the projection is given
  in \lc03.}. Following \lc03{} we indicate with $\Sigma(\ell)$ the
projected (mass or light) density distribution, where
\begin{equation}
\cases{
\ell^2\equiv\xp^2 +\yp^2/\qt^2, \cr
\qt^2\equiv\cos^2\theta +q^2\sin^2\theta .\cr}
\label{eq:ABqt}
\end{equation}
The quantity $\ell$ determines the size of the elliptic isophotes, and
their \emph{ellipticity} is given by $\varepsilon =1-\qt$. For fixed
$\ell$ the major and minor isophotal semi-axes are $a=\ell$ and
$b=\qt\ell$, and the associated {\it circularized radius} is defined as
$\pi R_\ell^2=\pi ab$, i.e., $R_\ell=\sqrt{\qt}\ell$.  In particular,
the {\it circularized effective radius} is
\begin{equation}
\Re =\sqrt{\qt}\ellef, 
\label{eq:recirc}
\end{equation}
where $M_{\rm p} (\ellef)=M/2$, with $M_{\rm p} (\ell)$ being the
projected mass within $\ell$. Thus the identity $\Re =\sqrt{\qt}\Reo$
(where $\Reo$ the effective radius of the model when seen face--on or in
case of spherical symmetry) is a common property of all axisymmetric
homeoidal distributions, independently of their specific density
profile, and so the dependence on $\theta$ of the circularized effective
radius of the present models and those in \lc03{} is identical.

The projected velocity fields at $(\xp,\yp)$ are obtained by numerical
integration of the projection along $\nv$ of their spatial counterpart.
This is done by transforming the corresponding spatial velocity moments
from cylindrical to Cartesian coordinates (\lc03, Eqs.~[21]-[27]). Note
that, in presence of a non-zero projected streaming velocity field
$\vp$, the velocity dispersion accessible to observation is
\begin{displaymath}
\sigobs(\xp ,\yp) \equiv \frac{1}{\Sigma (\ell)} \,
\int_{-\infty}^\infty \rho\,\,
\overline{\left(\langle\vv ,\nv\rangle - \vp\right)^2} d\zp =
\end{displaymath}
\begin{equation}
\:\;\;\quad\quad\quad\quad=\sigps+\Vps-\vp^2,
\label{eq:Sigsiglos}
\end{equation}
where ${\langle ,\rangle}$ is the standard inner product, and for the
definition of $\Vps$ (the \los{} integration of $\vn^{2}$, i.e.\ the
squared component along $\nv$ of the streaming velocity field) and of
$\sigps$ we again refer to \lc03{}. Note also that, independently of the
\los orientation, $\vn = 0$ on the isophotal minor axis $\yp$, and thus
$\sigobs =\sigps$ there.  In addition, $\sigobs =\sigps$ everywhere when
observing the galaxy face-on ($\theta =0$), or in the case $k=0$ (no
rotation).  Finally, since the observed velocity dispersion is always
measured within the aperture of the spectrograph, we integrate
$\sigob^2$ over circular apertures of radius $\rap$:
\begin{equation}
  \label{eq:sigobsa}
M_{\rm p}(\rap)\sigap^2(\rap) \equiv \int_{R' \leq \rap} \Sigma (\xp, \yp)
\sigobs(\xp,\yp)\,d\xp\, d\yp.
\end{equation}
In the following we will identify the quantity $\sigc$ appearing in the
FP with $\sigap$.

\subsection{The numerical code}
\label{sec:numcode}

The solution of the Jeans equations and the projection of the density
and of the various kinematical quantities are calculated with a
grid-based numerical code developed by one of us (A.R.). The gravitational
potential is obtained by solving the Poisson equation $\nabla^2 \phi = 4
\pi G \rho$ in spherical coordinates $(r, \vartheta, \varphi)$,
specializing the Londrillo \& Messina (1990) spectral method to
axisymmetric systems. The $\vartheta$ (co-latitude) dependence of $\rho$
and $\phi$ is described with standard Legendre polynomials, while the
spherical radius is mapped as
\begin{equation}
  \label{eq:remapping} 
  r \equiv \tan^{2} \left( u / 2 \right), \ \ \ \ \ (0
  \leq u \leq \pi).
\end{equation}
The resulting radial $u$-Laplace operator, when acting on Chebyshev
polynomials $T_n(\cos u)$, originates a finite linear combination of
such orthogonal functions, used to expand the density and the
gravitational potential.  Equations (\ref{eq:jeans1}) and
(\ref{eq:jeans2}) are then integrated with a cubic spline algorithm on a
logarithmic $(R,\, z)$ grid, in order to sample the model central
regions at high resolution (as required by the density cusp and by the
possible presence of a SMBH). Since the projection coordinate system is
tilted with respect to the natural coordinate system by an angle
$\theta$, the grid points of the two systems in general do not coincide,
and so an interpolation and the \los{} integrations are performed with a
cubic spline algorithm. For obvious symmetry reasons, the computations
are restricted to the first quadrant.

The double precision C code is organized in a library of functions, and
all the relevant intrinsic and projected fields of each model are
computed in $\sim 100 \, \mbox{s}$ on a $1.33 \mbox{GHz}$ processor, for
a $512^3$ grid.  The code accuracy has been tested by solving the
Poisson, Jeans, and projected equations for the Miyamoto-Nagai (1975)
models (see also Ciotti \& Pellegrini 1996), for the spherical
$\gamma$-models, and for the Ferrers ellipsoids: for these models all
the relevant dynamical properties and, to some extent, even the
projected fields are known analytically. In addition, we also compared
our numerical results with the analytical (asymptotic) models presented
in the Appendix and in \cb05{}, obtaining relative errors~$\la 10^{-3}$
for the intrinsic dynamical properties and~$\lsim 10^{-2}$ for the
projected fields (also in the central regions and in presence of a
SMBH). Finally, for all models presented in the following section we
verify numerically the projected virial theorem
\begin{equation}
  \label{eq:proj-virial-th}
  2 n_i n_j K_{ij} = -n^2_1 W^2_{11} -n^2_2 W^2_{22}
  -n^2_3 W^2_{33},  
\end{equation}
with $K_{ij}$ and $W_{ij}$ being the kinetic and potential energy
tensors (e.g. Ciotti 2000, see also Eqs.~[31], [A.8], [A.11], [A.12] in
\lc03).

\section{General results}
\label{sec:results}

\begin{figure*}[htbp]
  \centering
  \includegraphics[width=17cm]{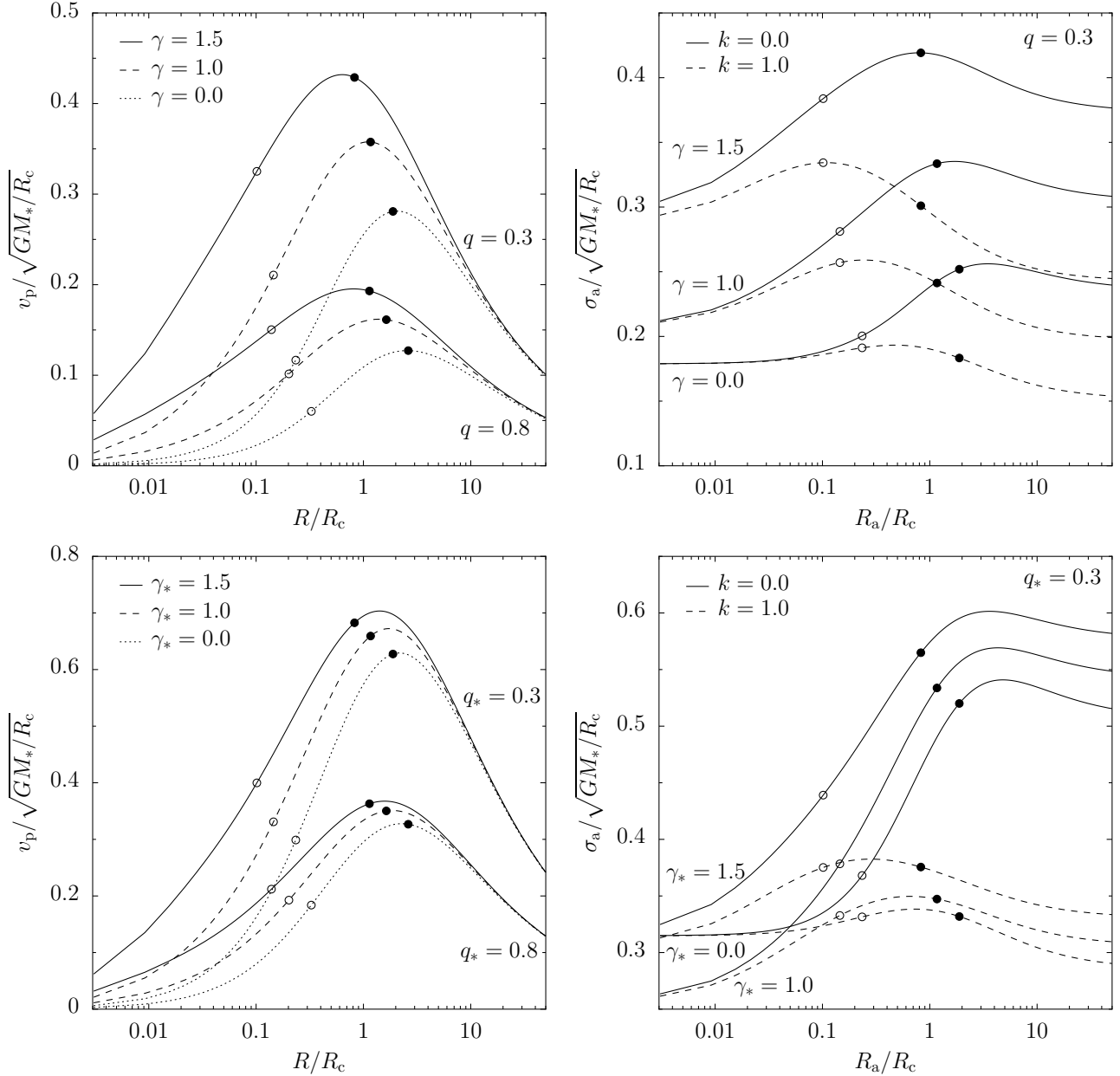}
  \caption{Edge-on, major axis projected streaming velocity $\vp$ (left
    panels), and aperture velocity dispersion $\sigap$ (right panels)
    for one (upper panels) and two-component (lower panels) galaxy
    models. Projected streaming velocity is shown for isotropic models
    only, while $\sigap$ for isotropic and fully tangentially
    anisotropic (i.e.\ not rotating) models. Aperture radii
    corresponding to $\Re /8$ and $\Re$ are marked as empty dots
    ($\circ$) and solid dots ($\bullet$), respectively. In two-component
    models the dark matter halo is described by a spherical Hernquist
    (1990) model ($\gamma = 1$, $q = 1$ in Eq.~[\ref{eq:rho}]), with
    $M_{h} = 5 \, M_{\ast}$ and $R_{h} = 2 \, R_{c}$.}
  \label{fig:vrot-sigmaAp}
\end{figure*}

Before studying in detail the effect of ordered rotation on the measured
velocity dispersion in elliptical galaxies and its consequences on the FP, we
present a few representative models in order to illustrate the general
properties of their projected kinematical fields (see also \dg).  In 
particular, since ordered rotation enters in the definition of $\sigma^{2}$
through the quantities $\vp$ and $\Vp$ (Eq.~[\ref{eq:Sigsiglos}]
and~[\ref{eq:sigobsa}]), and since these quantities are maximized for
isotropic rotators seen edge-on, we restrict to this configuration in the
following discussion.

In Fig.~\ref{fig:vrot-sigmaAp} (top left panel) we show $\vp$ along the major
axis for three different values of the galaxy density inner slope ($\gamma =
0, 1, 1.5)$ and two different flattenings $(q = 0.3, 0.8)$ in one-component
isotropic models without central SMBH.  Consistently with
Eqs.~(\ref{eq:hom-delta-sigma}) and~(\ref{eq:vphi}), and with
\cb05{}~(Eq.~[C.8]), in central regions $\vp$ vanishes in all the considered
cases. The projected streaming velocity also vanishes independently of
$\gamma$, as $R \rightarrow \infty$, where the density profile $\propto
m^{-4}$. The major axis projected streaming velocity thus presents a
rapid increase up to a maximum, placed very near to the model circularized
effective radius, followed by a mild decrease for increasing radial
coordinate (note the logarithmic scale of the x-axis). Apparently, at fixed
galaxy mass and scale-length, the main parameters determining the projected
velocity $\vp$ are the galaxy flattening $q$, and the slope $\gamma$ of the
central cusp. In Fig.~\ref{fig:vrot-sigmaAp} (top right panel) we show the
aperture velocity dispersion $\sigap$ for the same models (with $q = 0.3$) in
the left panel, either supported by tangential anisotropy ($k = 0$, solid
line), or by ordered rotation ($k = 1$, dashed line). In general, $\sigap$ is
larger for $k = 0$ than for $k = 1$ at any fixed aperture, and this is so
because tangential anisotropy is maximal when $k=0$, while the streaming
motions (that are maximal for $k = 1$) contribute to $\sigob$ with the term
$\Vp^2-\vp^2$ which is small (see Eq.~[\ref{eq:Sigsiglos}]). This dependence
of $\sigap$ on $k$ is explicit in Eq.~(29) in \cb05{}. In any case, also
$\sigap$ presents a maximum, which is near $\Re/8$ for the isotropic
rotators, and near $\Re$ for non-rotating models.

An important quantitative result of our models is that $\sigap$ differs
less than 15\% between the non-rotating and maximally rotating cases,
when $\rap \lsim \Re / 8$ since $\vp$ and $\Vp$ vanish near the galaxy
center.  When integrating $\sigob$ over a larger area $(\rap\sim \Re)$,
however, the differences between $\sigap$ of the anisotropy-supported
and the rotation-supported models increase for increasing $\gamma$,
attaining a value of about 30\% in the most extreme case $(\gamma =
1.5)$.

In Fig.~\ref{fig:vrot-sigmaAp} (bottom panels) we show the same quantities of
the corresponding upper panels, for identical galaxy models at which we added
a spherically symmetric ($q_{h} = 1$) dark matter halo with $M_{h} = 5
M_{\ast}$ and $R_{h} = 2 R_{c}$. For the halo central cusp we adopted $\gamma
= 1$, in order to mimic the central behavior of the NFW density
profile. Qualitatively the results are similar to those of the one-component
models, the main effect being the substantial increase of $\sigap$ for large
apertures in the $k = 0$ case. Note also that in the central regions the
effect of the adopted dark matter halo is negligible, as revealed by
Eqs.~(\ref{eq:hom-sigma-R})-(\ref{eq:hom-delta-sigma}) with $\MBH = 0$.
Analogous results are obtained also with halos with different mass or scale
radius.

Compared to the observations (e.g., Mehlert et al. 2000), while the
model rotational velocity profiles are quite realistic, the aperture velocity
dispersion presents a central dip that seems at odd with what empirically
inferred from the data (e.g., J{\o}rgensen et al. 1995; Mehlert et al. 2003;
Cappellari et al. 2005). This is an unrealistic, but very well known, feature
of several dynamical models with cuspy density distributions, which produce
centrally vanishing velocity dispersion profiles (see Eq.[A.4]; e.g., Bailey
\& MacDonald 1981; Dehnen 1993; Tremaine et al. 1994; and Bertin et al. 2002
for a general discussion in the case of spherical symmetry).  Such a central
dip in the models often occurs at small radii, so that it is undetectable in
the data or it nicely agrees with few observed cases (e.g., Graham et
al. 1998; Emsellem et al. 2004; Cappellari et al. 2005). This is also what we
find for our isotropic rotators, but cannot apply to the case of non-rotating
models, which show a peak in the \los velocity dispersion profile well far
from the galaxy center, near the effective radius. However, we stress again
that here we are considering the most extreme (thus also possibly
unrealistic) cases, in order to maximize the effects of ordered rotation,
while these unobserved features become much weaker for milder flattenings,
intermediate \los inclinations, and non-zero rotational support. We therefore
consider our results reliable and conclude that the contribution of the
rotational velocity to the \emph{observed} central velocity dispersion ($\rap
\lsim \Re / 8$) is usually negligible for elliptical galaxies in the local
Universe, also in presence of a cuspy dark matter halo. This implies, that
\emph{a systematic increase of rotational support with decreasing galaxy
luminosity is not at the origin of the tilt of the FP of elliptical galaxies}
(see also Busarello et al.\ 1997, \lc03{}). However, our results also suggest
that particular care should be used when constructing the FP of galaxies at
high redshift (where spectroscopic apertures usually enclose a large fraction
of the galaxy).  In fact, the increasing difference of $\sigap$ between
rotating and non rotating galaxies, with increasing $\rap$ (dashed line in
Fig.~\ref{fig:vrot-sigmaAp}) would lead, if not properly taken into account,
to underestimate the galaxy mass (when interpreting $\sigap$ by means of
virial estimators based on spherical, not rotating models) up to 70\% for
low-mass and rotating objects (see also Bender, Burstein \& Faber 1992, van
Albada, Bertin \& Stiavelli 1995).  These aperture effects might so
contribute (in part) to the observed decrease of the mass-to-light ratio of
low-mass galaxies at high redshift (e.g., di Serego Alighieri et al.\ 2005;
Treu et al.\ 2005; van der Wel et al.\ 2005).

\begin{figure*}[htbp]
  \centering{}
  \includegraphics[width=17cm]{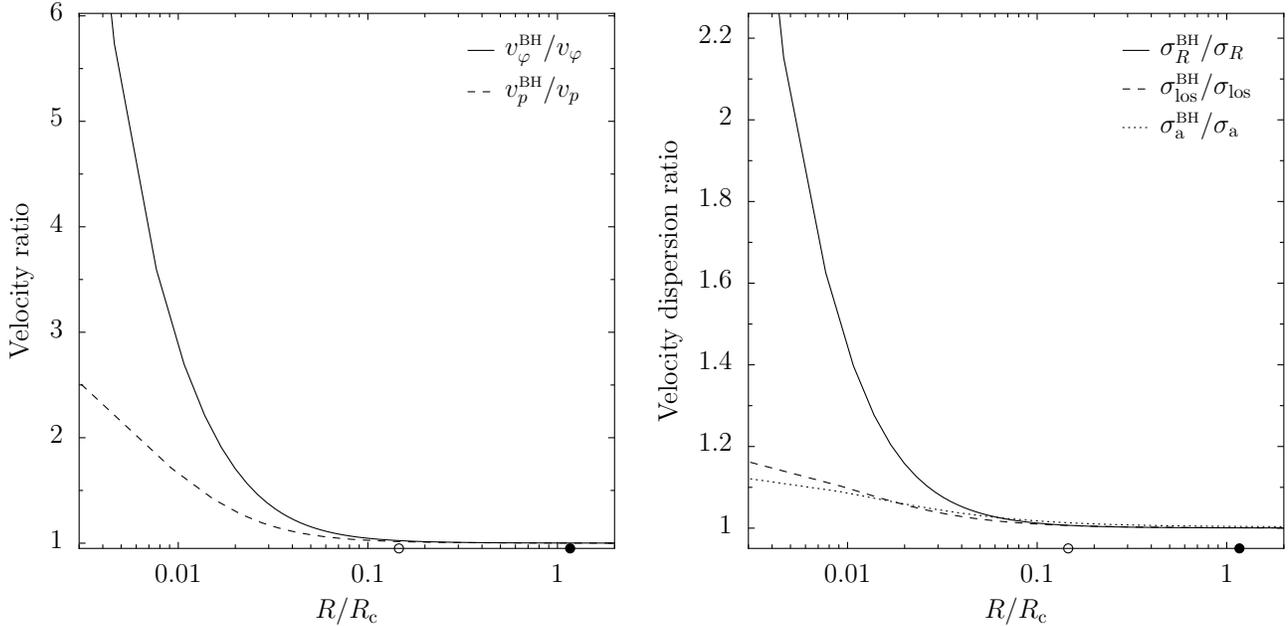}
  \caption[]{{\it Left panel}: Intrinsic (solid line) and
    projected (dashed line) streaming velocities on the equatorial plane
    ($z = 0$) of a models with $\gamma = 1$, $q = 0.3$, $k = 1$, and a
    SMBH with $\MBH = 10^{-3}\, M_{\ast}$ normalized to the
    corresponding quantity of the same model without the SMBH.
    \emph{Right panel}: Intrinsic (solid line), line-of-sight (dashed
    line), and aperture velocity dispersion ratios (dotted line) for the
    same model. The aperture velocity dispersion is calculated for a
    circular aperture $R_{a} = R$. The empty ($\circ$) and solid
    ($\bullet$) dots correspond to $\Re / 8$ and $\Re$, respectively.}
\label{fig:allvel-bh}
\end{figure*}

\subsection{Effects of a central super-massive black hole}
\label{sec:effects-bh}

It is now commonly accepted that SMBHs reside in the center of stellar
spheroids, and that their mass $\MBH$ scales almost linearly with the
stellar mass of the parent galaxy, with $\MBH / M_\ast \approx 10^{-3}$
(Magorrian et al. 1998). An order-of-magnitude estimate of the radius
$\rBH$ of the ``sphere of influence'' (i.e., the region within which the
gravitational effects of the SMBH on the stellar orbits are significant)
is usually obtained as $G \MBH /\rBH = 3\,\sigma_\ast^2/2 \approx G
M_\ast/(2\,r_{\rm vir})$, i.e., $\rBH/\Re \approx 6\times 10^{-3}$
(where $\sigma_\ast$ is the one-dimensional stellar velocity dispersion,
and $r_{\rm vir}/\Re \simeq 2.97$ for a $R^{1/4}$ galaxy; see Ciotti
1991).  Thus, SMBH effects should be detectable on scales of few parsecs
or tens of parsecs only, in any case well within the standard central
aperture of $\Re/8$. However, while in galaxy models without SMBH
$\vcphi$ and $\vp$ vanish at the center for $\gamma < 2$, in presence of
a SMBH $\vcphi^{\rm {\footnotesize BH}} \propto \vp^{\rm {\footnotesize
    BH}} \propto 1/\sqrt{R}$, and in principle even an unresolved but
luminosity weighted, central kinematical spike could produce detectable
effects on $\sigap$.  The results obtained with our models are
illustrated in Fig.~\ref{fig:allvel-bh} (left panel), where we show the
spatial and projected streaming velocity along the major axis of an
isotropic $\gamma=1$ model with $q=0.3$ and with a central SMBH having
$\MBH / M_\ast =10^{-3}$. It can be seen that while $\vcphi$ increases
by nearly one order of magnitude in the innermost regions, $\vp$ (the
only one directly accessible to observations) is much less affected by
the presence of the SMBH. Note that from Eq.~(\ref{eq:hom-delta-sigma}),
and~\cb05{}~(Eq.~[C.8]), $\vcphi^{\rm {\footnotesize BH}}/\vcphi\propto
\vp^{\rm {\footnotesize BH}}/\vp\propto R^{(\gamma-3)/2}$. In all
explored models we found that the SMBH influence is negligible for $R
\gsim \Re / 8$.

In Fig.~\ref{fig:allvel-bh} (right panel) we show for the same model the
intrinsic one-dimensional and \los{} velocity dispersions along the
major axis, and the aperture velocity dispersion as a function of
$\rap$. Note that from Eq.~(\ref{eq:hom-sigma-R}) and Eqs.~(C.3),
(29)-(30) of \cb05{} we expect $\sigR^{\rm {\footnotesize
    BH}}/\sigR\propto \sigob^{\rm {\footnotesize BH}}/\sigob \propto
\sigap^{\rm {\footnotesize BH}}/\sigap\propto R^{(\gamma - 3)/2}$. In
fact, our numerical results confirm this radial trend, and also show
that the SMBH influence decreases along the sequence $\sigma_z$,
$\sigma_{\rm los}$ and $\sigap$ at fixed radial coordinate. A comparison
of the two panels in Fig.~\ref{fig:allvel-bh} shows that velocity
dispersions are less affected than rotational velocities by the presence
of a central SMBH.  We thus conclude, that, in general, the presence of
a SMBH can be neglected when using the measured ``central'' ($\rap
\simeq \Re / 8$) velocity dispersion, for the construction of the
Faber-Jackson (1976), FP, and $\MBH$-$\sigma$ relations (Ferrarese \&
Merritt 2000, Gebhardt et al.\ 2000), as well as the
$v/\sigma$-ellipticity plane (Illingworth 1977, Binney 1978).
Accordingly, in the following Sections we will not consider a central
black hole in the models.

\section{Contribution of rotational velocity to the FP thickness}
\label{sec:fp-thickness}

\begin{figure*}[htbp]
  \centering{}
  \includegraphics[width=17cm]{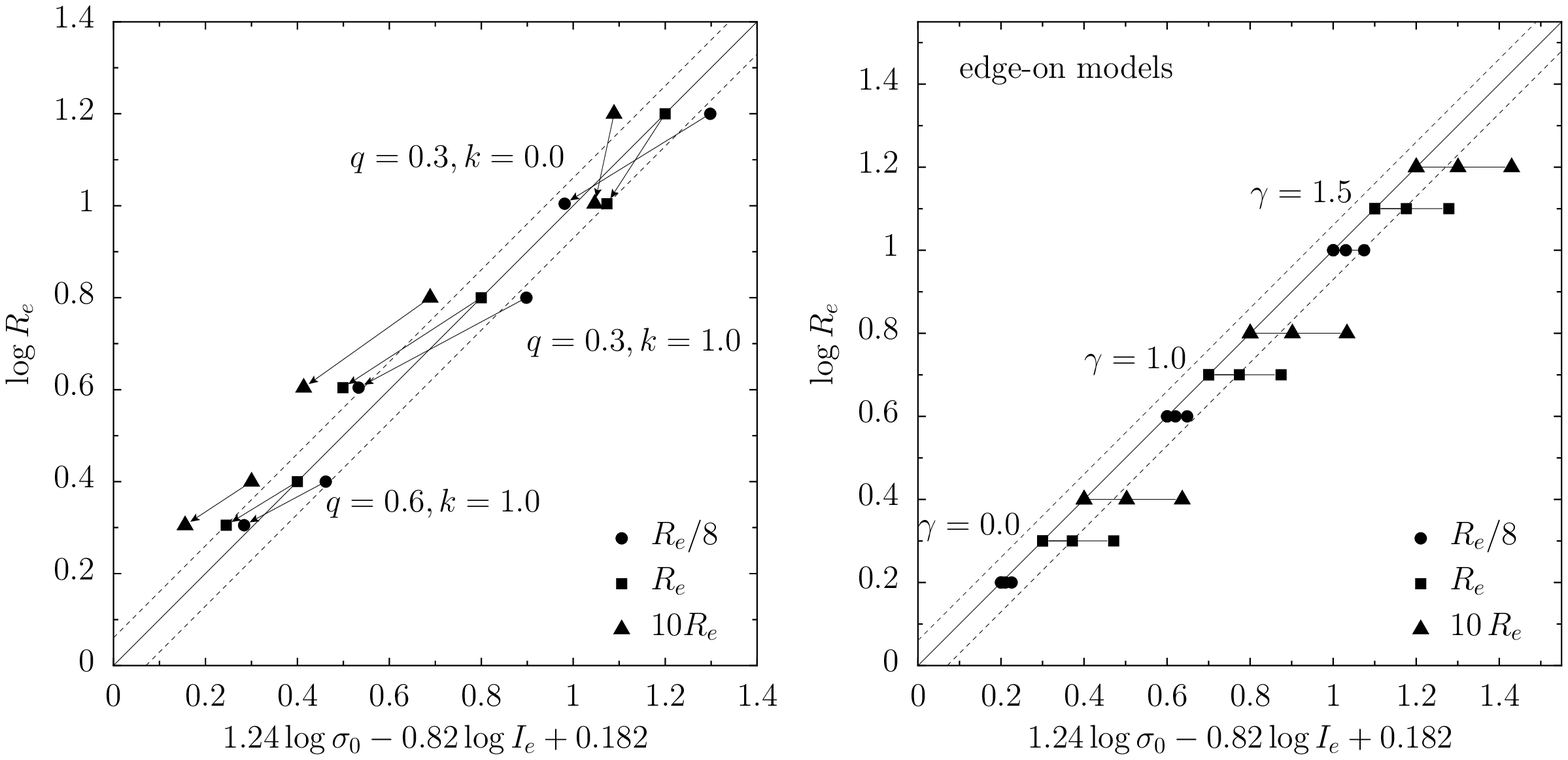}
  \caption[]{\emph{Left panel}: Projection effects on the galaxy models
    in the space where the FP for Coma cluster galaxies is seen edge-on
    (solid line) with 1-$\rms$ scatter (dashed line) from JFK.  The
    intrinsic model parameters are given by labels, and $\gamma = 1$ for
    all models.  For each model three spectroscopic apertures for
    $\sigap = \sigc$ are considered ($\Re / 8$: dots, $\Re$: square, $10
    \Re$: triangles). The \los inclination increases from $\theta = 0$
    (face-on) to $\theta = \pi / 2$ (edge-on) along the arrows, and the
    cases $\theta = 0$, $\rap = \Re$ are \emph{arbitrarily} placed on
    the FP best-fit line. \emph{Right panel}: Maximum rotational effects
    as a function of intrinsic flattening and spectroscopic aperture
    $\rap$ for three edge-on models ($\gamma = 0, \, 1, \, 1.5$). The
    isotropic rotators with $q = 0.6$ are \emph{arbitrarily} placed on
    the FP while, on each horizontal line, the middle points correspond
    to models with $k=0$ and $q = 0.6$, while the rightmost points to
    models with $k=0$ and $q = 0.3$.}
\label{fig:fp-edge-on-view}
\end{figure*}

In a follow-up of \lc03{} we now investigate the contribution of
projection and ordered rotation on the observed FP thickness. As
anticipated in the Introduction, the only differences with respect to
the results of \lc03{} are possibly due to the new value of $\sigc$,
since the variations of the circularized effective radius $\Re$ (and of
$\Ie$) are independent of the specific stellar density profile adopted
(provided it is stratified on similar ellipsoids). We start by
illustrating how models ``move'' in the edge-on view of the FP as a
function of their intrinsic ($\gamma$, $q$, $k$) and observational
($\theta$) parameters.  In Fig.~\ref{fig:fp-edge-on-view} (left panel)
we show the edge-on view of the FP of Coma cluster ellipticals as given
by Eq.~(\ref{eq:FP}) (solid line), and its $\rms$ scatter $\simeq 0.057$
(dashed line, as obtained by JFK for galaxies with $\sigc\ge 100$ km/s,
and after correction for measurement errors). The three families of
models superimposed are determined (from top to bottom) by $(q,\, k) =
(0.3,\, 0)$, $(0.3,\, 1)$, and $(0.6,\, 1)$. For simplicity, $\gamma =
1$ in all models, while three different aperture radii $\rap$ are
considered to measure $\sigc = \sigap$, namely $\rap = \Re/8$ (dots),
$\Re$ (squares), $10 \, \Re$ (triangles).  The cases corresponding to
$\theta = 0$ and $\rap = \Re$ are arbitrarily placed on the FP best-fit
line, and the displacements correspond to $\theta$ increasing from $0$
to $\pi/2$.  According to Eq.~(\ref{eq:ABqt}), when the \emph{los}
inclination changes from $0$ to $\pi/2$, $\Re$ decreases (and $\Ie$
increases), thus producing a vertical down-shifts towards the left of
the representative models; as expected, displacements are smaller for
rounder systems. The additional effect of rotational velocity produces
horizontal displacements which depend mainly on the adopted $\rap$. For
$k = 1$ (isotropic rotators), the variation of $\sigap$ with the viewing
angle $\theta$ is only slightly dependent on the considered
spectroscopic aperture, so that the model displacements are almost
parallel to each other, but not to the FP. On the contrary, the
displacements of non-rotating anisotropic galaxies change considerably
both in magnitude and direction as function of $\rap$. We have verified
that these results are almost independent of the model central cusp
$\gamma$. Thus, \emph{due to projection effects, galaxies move in
  directions which are not exactly parallel to the edge-on FP, but the
  entity of these displacements is small enough to always maintain the
  models within the observed FP thickness}.  A comparison with Fig.~5 in
\lc03 reveals remarkably similar behaviors.

We now restrict the analysis to rotational effects as a function of
density slope and galaxy flattening, and in
Fig.~\ref{fig:fp-edge-on-view} (right panel) we plot a selection of
representative models. As in the left panel, we consider the edge-on
view of the FP, and we describe how models ``move'' when varying the
relative amount of rotational and anisotropy support. In particular we
show the behavior of three families of models seen edge-on, with
different central cusp slopes ($\gamma = 0, \, 1, \, 1.5$), and with
different flattening $q$.  Isotropic models ($k = 1$) are arbitrarily
placed on the FP best-fit line. A reduction of the amount of rotational
support to zero ``moves'' models toward the right, placing them at the
intermediate points, corresponding to $k = 0$ and $q = 0.6$.  More
flattened models move even further, until the rightmost points, that
represent models with $k = 0$ and $q = 0.3$.  These displacements are
due to the increasing values of the central velocity dispersion $\sigc$
as the rotational support becomes negligible, and their amount increases
with both the aperture radius and the galaxy flattening, while it is
almost independent of the shape of the central cusp $\gamma$, in
agreement with Fig.~\ref{fig:vrot-sigmaAp}. Note that a preliminary
analysis of this problem was given by \cb05{}, who adopted homeoidal
expansion to investigate the behavior of the projected velocity
dispersion in cuspy oblate models with a central SMBH. A comparison of
our results with Fig.~2 of \cb05{} and results discussed therein,
reveals very similar behaviors. Note that the models showed in
Fig.~\ref{fig:fp-edge-on-view} (right panel) are all observed edge-on,
in order to maximize the rotation effect; when all the possible
line-of-sight and the distribution of intrinsic ellipticities are taken
into account (as in the Monte-Carlo approach in \lc03{}) the statistical
displacements would be essentially negligible.

From the study of the models presented in this Section, and from their
comparison with models in \lc03{} and \cb05{} we then confirm and extend
the conclusions of \lc03{} to cuspy galaxies, \emph{i.e.\ that
  projection effects only marginally contribute to the FP thickness,
  \emph{90\%} of it being due to variations, from galaxy to galaxy, of
  their intrinsic physical properties}. In particular, we proved that
variations of rotational support from galaxy to galaxy \emph{only
  marginally contribute to the FP thickness}.

\section{The $v/\sigma$-ellipticity plane}
\label{sec:vsigma-epsilon-plane}
\begin{figure}[tbp]
\resizebox{0.95\hsize}{!}{\includegraphics[]{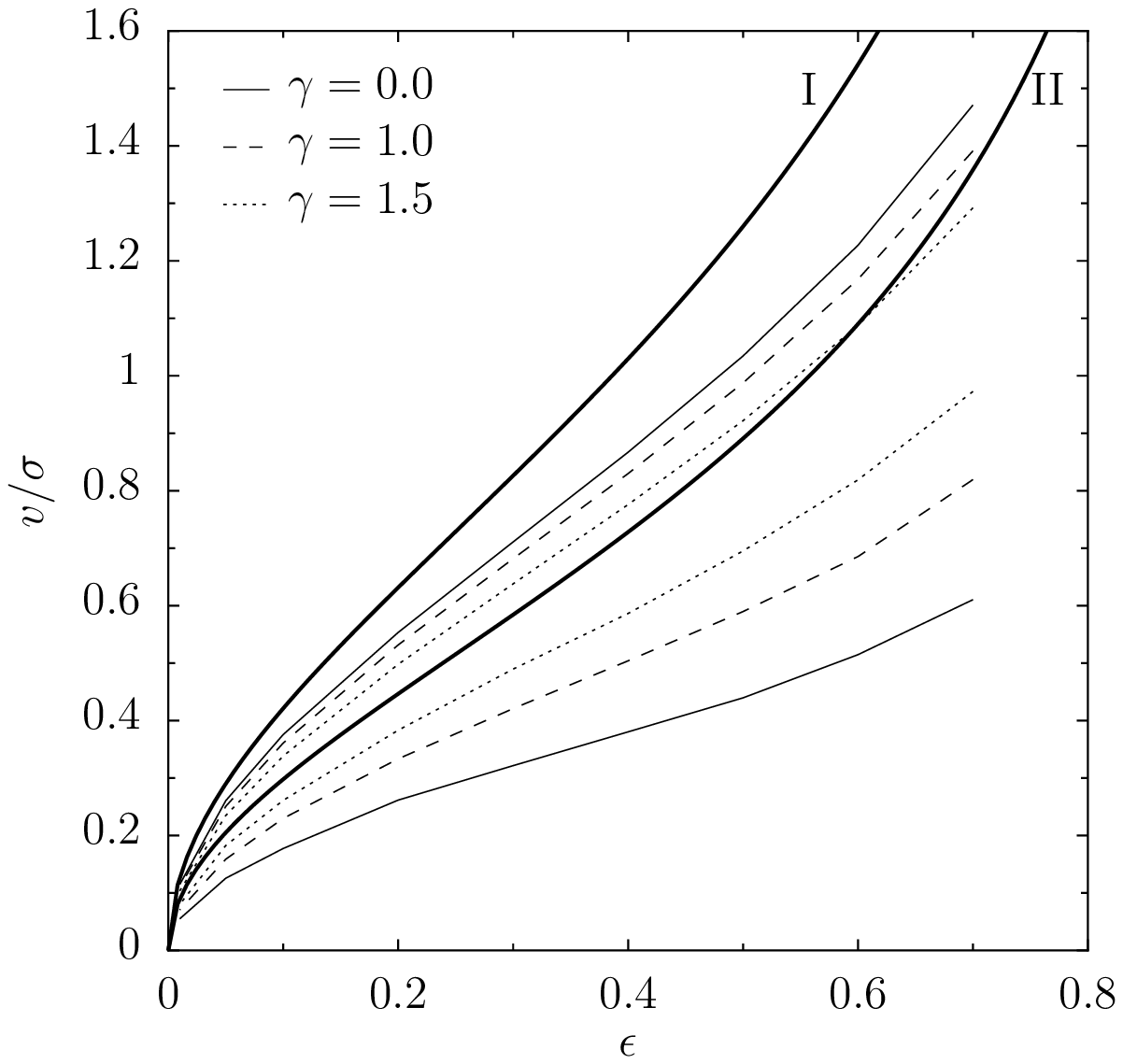}}
\caption{Rotational parameter for isotropic, edge-on models as a
  function of their ellipticity. The solid thick curve I is the locus of
  isotropic, classical ellipsoids as obtained from the tensor virial
  theorem (Binney 1978; see also Eq.[4.95] in BT), while solid line II
  is derived from the sky-averaged tensor virial theorem (Eq.~[26] of
  Binney 2005, with $\delta = 0$ and $\alpha = 0$) . The upper set of
  curves represents our models when $\vp = v_{{\rm p},\mathrm{max}}$,
  while the lower set of curves represents our models when $v =
  \vp(\Re/8)$. In all cases we adopt $\sigma = \sigap(\Re / 8)$.}
\label{fig:vsigma}
\end{figure}

With the aid of the developed models we finally study how the
$v/\sigma$-ellipticity plane (an important tool used to investigate to
what extent the galaxy flattening is due to velocity dispersion
anisotropy, e.g.\ see Illingworth 1977, Binney 1978) is populated by
galaxies as a function of the adopted aperture $\rap$ and of their
streaming velocity support. Traditionally, observational data are the
peak projected streaming velocity $v_{\rm p,max}$ a ``central'' aperture
velocity dispersion $\sigc$, and galaxy ellipticity. Data points are
then compared with the family of curves obtained from the tensor virial
theorem, where $v^{2}$ and $\sigma^{2}$ are the mass weighted square
streaming velocity and velocity dispersion computed over the whole
galaxy. As well known, the tensor virial theorem predictions are
independent of the specific galaxy density profile as far as it is
homeoidally stratified (Roberts 1962). For example, the thick solid line
I in Fig.~\ref{fig:vsigma} is obtained from Eq.~(4.95) of BT (see also
Eq.~[9] of Binney 2005), and corresponds to isotropic rotators seen
edge-on.  Galaxies whose representative points lie significantly below
this curve are considered anisotropy supported.  Unfortunately it is not
straightforward to connect the virial quantities to the observable ones.
For this reason, the tensor virial theorem curves are often
``corrected'' to take into account projection effects, and in general
these corrections makes curve I flatter (e.g., see Eq.~[4.5] and
Figs.~[8]-[9] in Evans \& de Zeeuw 1994, and Figs.~[4.5]-[4.6] in BT).
The situation became even more complicate when Evans \& de Zeeuw (1994)
showed that $v/\sigma$ for their nearly isotropic ``power-law'' galaxy
models, not only was systematically below curve I of classical spheroids
(and significantly so at large ellipticities), but also below the
``corrected'' locus. Of course, this behavior is not unexpected because,
as discussed by Evans \& de~Zeeuw (1994), the density of their model is
not stratified on ellipsoidal surfaces. In addition, their projected
streaming velocity does not have a maximum, and its fiducial value must
therefore be taken at some arbitrary distance from the center.
Interestingly, a very similar behavior is also shown by the isotropic
models described in \cb05{} (Fig.~[2]): in this case, due to the models
scale-free nature, $\vp$ and $\sigap$ adopted to construct $v/\sigma$
were taken at the same radius.

Fortunately, it is now possible to observationally map the rotation
speed and the velocity dispersion over a substantial fraction of a
galaxy's image (e.g.\ the SAURON project, see de Zeeuw et al.\ 2002,
Cappellari et al.\ 2005). Thus, one can use the sky-averaged quantities
to define $v/\sigma$, and predict its trend from the projected virial
theorem (e.g.\ Binney 2005, see also Ciotti 1994, 2000). Curve II in
Fig.~\ref{fig:vsigma} shows the corresponding locus of the edge-on
isotropic rotators, as given by Binney (2005).

Here we use our (isotropic, edge-on, and homeoidally stratified) models
to study the effects of light profile central cusps and different
definitions of the projected streaming velocity, in determining the
model position in the $v/\sigma$-$\epsilon$ plane, while for simplicity
we fix $\sigma = \sigap (\Re / 8)$. In particular, we consider the cases
$v = v_{\rm p,max}$ (the maximum projected velocity) and $v = v_{\rm p}
(\Re / 8)$, as representative of the measured rotation.  The resulting
trends are shown in Fig.~\ref{fig:vsigma}, and it is clear how \emph{all
  the resulting curves lie below the (uncorrected) locus of isotropic,
  classical ellipsoids obtained from the tensor virial theorem (solid
  line \emph{I})}.  Deviations with respect to the ``classical''
expectation are larger when $v = v_{\rm p} (\Re / 8)$, a case analogous
to those considered by \cb05{}.  Remarkably, our analysis shows that
\emph{curve \emph{II (derived from the sky-averaged virial theorem)}
  provides a much better estimate of $v/\sigma$ (and so, of anisotropy)
  than the original one based on the (unprojected) tensor virial
  theorem, when $\sigma = \sigap (\Re / 8)$ and $v = v_{\rm p,max}$}.
Note also how the central cusp slope can be important, especially when
using small aperture to determine $v$.

Thus we conclude that particular care should be placed on the choice of the
theoretical $v/\sigma$ locus to be used for comparison with the data, as a
function of the quality and extension of the data themselves (see also
\dg).

\section{Discussion and conclusions}
\label{sec:disc-concl}
In this paper we have quantified the contribution of ordered rotation to
the FP tilt and thickness. We adopted a class of oblate galaxy models
with central cusp and adjustable flattening and amount of rotational
support. We also considered models with a dark matter halo and with a
central super-massive black hole. The associated Jeans equations have
been (numerically) solved under the assumption of an underlying
two-integrals distribution function. The main results can be summarized
as follow:
\begin{enumerate}
  
\item For the adopted family of models the effect of a central SMBH is
  stronger on streaming velocities than on velocity dispersions.
  However, for mass ratios of the order of those predicted by the
  Magorrian relation, the effect is in general negligible, and the
  presence of a central SMBH can be neglected when studying relations
  such as the FP.
  
\item In agreement with previous results of \lc03{} and \cb05{}, also
  for the adopted class of cuspy models the contribution of streaming
  motions to the observed velocity dispersion is negligible when
  small/medium apertures ($\lsim\,\Re$) are used for the spectroscopic
  observations. This implies that a systematic decrease of rotational
  support with increasing luminosity {\it is not} at the origin of the
  tilt of the FP of elliptical galaxies at low-redshift.
  
\item In general, the aperture velocity dispersion measured at $\rap
  \gsim \Re$ is larger for not rotating models than for rotationally
  supported models (and the difference increases significantly when a
  dark matter halo is present). This must be taken properly into account
  when studying the FP, or estimating the mass-to-light ratios at high
  redshift.
    
\item When observed from different \los inclinations, models move in the
  $(\log \Re, \, \log I_{e}, \, \log \sigc)$ space along directions that
  are not parallel to the edge-on view of the FP, thus confirming that
  projection effects do contribute to the observed FP scatter.  The
  models displacement depends mainly on their intrinsic flattening
  (being larger for more flattened systems), on the amount of rotational
  support, and on the aperture used to measure $\sigc$. These effects
  however, when weighted over the distribution of ellipticities and
  \los{} inclinations, would produce statistical displacements well
  within the FP thickness, in agreement with the conclusions of \lc03{}.
  
\item Finally, as a by-product of the present investigation we studied
  the position of the models in the $v/\sigma$-ellipticity plane, a tool
  used to discriminate between rotationally and velocity dispersion
  supported galaxies. We found that, while our models remain well below
  the (uncorrected) locus of classical ellipsoids as determined from the
  tensor virial theorem, their position is well predicted by
  sky-averaged projected virial theorem (Binney 2005) if the peak
  rotational velocity is used.
\end{enumerate}

\begin{acknowledgements}
  We thank Pasquale Londrillo for advice on the numerical code, and the
  anonymous referee for useful comments. B.L. is supported by a INAF
  post-doctoral fellowship, and L.C. by the Italian Cofin ``Collective
  phenomena in the dynamics of galaxies''.
\end{acknowledgements}

\appendix
\section{A simple, analytical, cuspy and axisymmetric galaxy model with dark
  matter halo and central SMBH}
\label{sec:axisymm-cuspy-analyt}

We present here the internal kinematical fields of a family of power-law
galaxy models (constructed with a homeoidal expansion following \cb05)
with dark matter halo and central black hole. The seed distributions for
stars and dark matter are $\rho_\ast = \rho_{\ast 0}/m_\ast^\gamma$ and
$\rho_h=\rho_{h0}/m_h^\delta$, where $m^2_\ast = \tilde{R}^2 +
\tilde{z}^2/(1-\epsilon)^2$, $m^2_{\rm h} = \tilde{R}^2 +
\tilde{z}^2/(1-\eta)^2$ and, without loss of generality, the
scale-length of both distributions is $\rc$, with $\widetilde{R}\equiv
R/\rc$ and $\widetilde{z}\equiv z/\rc$. According to \cb05~(Eq.~[A.2]),
the (unconstrained) density and potential expansion are performed for
$\gamma \epsilon \leq 1$, $\delta \eta \leq 1$ (with $\gamma, \,\delta >
1$).  Thus, from \cb05~(Eq.~[26]), the model stellar density
distribution and the total potential are
\begin{eqnarray}
  \label{eq:hom-tot-rho}
  \widetilde{\rho}_\ast &=& \frac{1-\gamma\epsilon}{\tilde{r}^\gamma} + 
  {{\gamma\epsilon\tilde{R}^2}\over{\tilde{r}^{\gamma+2}}} \\
  \label{eq:hom-tot-pot}
  \widetilde\phi &=& 
  \widetilde{\phi}_\ast + \mathcal{R} \widetilde{\phi}_h -
  \frac{\mu}{\tilde{r}},
\end{eqnarray}
where the density is normalized to $\rho_{0 \ast}$, the gravitational
potentials to $4 \pi G \rho_{\ast 0} \rc^2$, $\mu \equiv \MBH / 4 \pi
\rho_{\ast 0} \rc^3$, and finally $\mathcal{R} \equiv \rho_{h0} /
\rho_{\ast 0}$.  The explicit expression for $\widetilde{\phi}_{\ast}$
and $\widetilde{\phi}_{h}$ is given in \cb05~(Eq.~[27]), and the model
(dimensionless) circular velocity is
\begin{eqnarray}
  \label{eq:hom-vc2}
  \widetilde{v}_c^2 &=&
  \frac{5 -\gamma -2\epsilon}{(5 -\gamma) \,(3 -\gamma) \,{\widetilde{R}}^{\gamma-2}} + 
  \nonumber \\
  && \mathcal{R}\,\frac{5 -\delta -2\eta}{(5 -\delta) \,(3 -\delta)
    \,{\widetilde{R}}^{\delta-2}} + 
  \frac{\mu}{{\widetilde{R}}}.
\end{eqnarray}
The Jeans equations for the stellar component are integrated following
\cb05[Eqs.(14)-(15)], retaining only first order terms in the
flattenings $\epsilon$ and $\eta$:
\begin{eqnarray}
\label{eq:hom-sigma-R}
\widetilde{\rho}_\ast \sigR^2 &=& \frac{\widetilde{r}^{2(1-\gamma)}}{(5 -\gamma)
    (3 -\gamma)} \left[\frac{5 -\gamma -4\epsilon}{2(\gamma -1)}  
     -\frac{(4-\gamma)\epsilon\,\widetilde{z}^2}{\widetilde{r}^2} \right] +
   \nonumber \\
    && \frac{\mathcal{R} \, \widetilde{r}^{2 -\gamma-\delta}}{(5 - \delta)(3 - \delta)} \times \nonumber \\
  && \quad \left[ \frac{(5 - \delta)(1 - \gamma \epsilon) - (4 - \delta)(\delta - 1)
    \eta}{\gamma + \delta - 2} \right. + \nonumber \\ 
    && \quad \qquad \left. \frac{(5 - \delta) \gamma \epsilon + (3 - \delta) \delta
    \eta}{\gamma + \delta} \, \frac{R^2}{\widetilde{r}^2} \right] + \nonumber \\
    && \frac{\mu}{(3+\gamma)\,\widetilde{r}^{1+\gamma}} \left(\frac{3 +\gamma
    -2\gamma\epsilon}{1 +\gamma} -\frac{\gamma \,\epsilon
    \,\widetilde{z}^2}{\widetilde{r}^2} 
    \right) 
\end{eqnarray}

\begin{eqnarray}
\label{eq:hom-delta-sigma}
\widetilde{\rho}_\ast \left(\overline{\tvphi^2} - \sigR^2\right) &=&
    \frac{2 \epsilon \,\widetilde{R}^2}{(5 -\gamma) \,(3 -\gamma)
    \,\widetilde{r}^{2\gamma}} + 
   \nonumber \\ 
   && \mathcal{R} \,\frac{2 \,\gamma \,\widetilde{R}^2}{\widetilde{r}^{\gamma
    +\delta}} \,\frac{\epsilon \,(5 - \delta) - \eta \,(3 - \delta)}{(5
    -\delta) \,(3 -\delta) \,(\gamma +\delta)} +
   \nonumber \\
  && \mu \,\frac{2 \,\gamma \,\epsilon \,\widetilde{R}^2}{(3 +\gamma)
    \,\widetilde{r}^{3+\gamma}} \, . 
\end{eqnarray}

\end{document}